\documentclass[%
 reprint,
superscriptaddress,
 amsmath,amssymb,
 aps,
]{revtex4-2}

\usepackage{graphicx}% Include figure files
\usepackage{dcolumn}% Align table columns on decimal point
\usepackage{bm}% bold math
\usepackage{hyperref}% add hypertext capabilities
\usepackage[T1]{fontenc}
\usepackage{braket}
\usepackage{color}

\graphicspath{{./fig/}}
\begin{document}

\preprint{NITEP 129}

\title{Effective polarization in proton-induced $\alpha$ knockout reactions}% Force line breaks with \\

\author{Tomoatsu Edagawa}
\email[Email address: ]{tomoatsu@rcnp.osaka-u.ac.jp}
\affiliation{Research Center for Nuclear Physics (RCNP), Osaka University,
Ibaraki Osaka 567-0047 Japan}%

 \author{Kazuki Yoshida}
 \affiliation{Advanced Science Research Center, Japan Atomic Energy Agency (JAEA), 
 Tokai, Ibaraki 319-1195, Japan}
%  \email{yoshida.kazuki@jaea.go.jp}

% \collaboration{MUSO Collaboration}%\noaffiliation

\author{Yoshiki Chazono}
\altaffiliation[Present address: ]{RIKEN Nishina Center for Accelerator-Based Science, 
2-1 Hirosawa, Wako 351-0198, Japan}
\affiliation{Research Center for Nuclear Physics (RCNP), Osaka University,
Ibaraki Osaka 567-0047 Japan}%
% \email{chazono@rcnp.osaka-u.ac.jp}%

\author{Kazuyuki Ogata}
\altaffiliation[Present address: ]{Department of Physics, Kyushu University, Fukuoka 819-0395, Japan}
\affiliation{Research Center for Nuclear Physics (RCNP), Osaka University,
Ibaraki Osaka 567-0047 Japan}
\affiliation{Nambu Yoichiro Institute of Theoretical and Experimental Physics (NITEP), Osaka City University, 
            Osaka 558-8585, Japan}%
% \email{kazuyuki@rcnp.osaka-u.ac.jp}%

% \collaboration{CLEO Collaboration}%\noaffiliation

\date{\today}% It is always \today, today,

\begin{abstract}
  The effective polarization of the residual nucleus in 
  the proton-induced $\alpha$ knockout reaction is investigated 
  within the distorted wave impulse approximation framework. 
  The strong absorption of the emitted $\alpha$ particle results in strong selectivity 
  on the reaction {\lq\lq}position'' depending on 
  the third component of the single-particle orbital angular momentum 
  of the $\alpha$ particle inside a nucleus, 
  hence on the spin direction of the reaction residue. 
  This is caused by a mechanism that is similar to the Maris effect, 
  the effective polarization in the proton-induced proton knockout reactions. 
  However, as a distinct feature of the effective polarization in 
  the $\alpha$ knockout process, the spin degrees of freedom of 
  the reacting particles play no role. 
  The $\alpha$ knockout process with complete kinematics can be 
  a useful polarization technique for the residual nucleus, 
  without actively controlling the spin of the proton.
\end{abstract}

\maketitle

%\tableofcontents

\section{INTRODUCTION}
Proton-induced nucleon knockout reactions, $(p,pN)$, have been developed to 
investigate the single-particle (s.p.) structure of nuclei~\cite{Jacob66,Jacob73}. 
In a recent review~\cite{Wakasa17}, the $(p,pN)$ reactions were shown to be 
a reliable spectroscopic tool, meaning that the deduced spectroscopic factors are 
consistent with those determined by electron-induced knockout reaction data, 
typically within a deviation of 15\%. In addition, the proton has the advantage of being able to knock neutrons out, whereas the electron can only strike charged particles. Then this has been opening a door to 
knockout reaction studies of neutron s.p.~levels as well as s.p.~structures of unstable nuclei. 
In recent years, a series of ($p,pN$) measurements has provided us with
new findings on the s.p.~structure and the magicity of unstable nuclei
~\cite{Olivier17,Kawase18,Elekes19,Taniuchi19,Chen19,Cortes20,Sun20,Lokotko20,
Tang20,Kubota20,Cortes20_2,ZYang21,Juhasz21,Juhasz21_V63,Browne21,Linh21}.
%Kubota20(dinutron)
On the reaction theory side, the quantum transfer-to-the continuum model (QTC)~\cite{Moro15,Gomes18}, the eikonal DWIA~\cite{Aumann13}, and the Faddeev--Alt-Grassberger-Sandhas theory (FAGS)~\cite{Cravo16,Crespo2019,Mecca2019} for knockout reactions have been developed; there are some benchmark type studies among them~\cite{Yoshida18b,Gomez20}.
The knockout reactions have also been utilized as a probe of nuclear clustering.
In Ref.~\cite{Tanaka21}, the first direct observation of $\alpha$ particle formation on Sn isotopes, triggered by the theoretical prediction by Typel~\cite{Typel14}, was confirmed by the $\alpha$ knockout reaction. 

It is shown that ($p,pN$) reactions mainly probe the surface region 
of nuclei because of the nuclear absorption~\cite{Aumann13,Gomes18,Aumann21}.
Recently, by choosing the kinematics of the $^{11}$Li($p,pn$) process, the correspondence between the correlation angle of the two neutrons and their {\lq\lq}location'' inside $^{11}$Li was investigated~\cite{Kubota20}; this work has triggered  intensive discussion from a theoretical point of view~\cite{Casal2021,Yamagami2022}. Even though the nuclear distortion tends to prohibit a simple interpretation of the measured cross section, the result of Ref.~\cite{Kubota20} implies that, when light nuclei are considered, ($p,pN$) may probe the rather interior region of them because of the relatively weak nuclear absorption.

On the other hand, the surface sensitivity is expected to be more emphasized in proton-induced
$\alpha$ knockout reactions, ($p,p\alpha$)~\cite{Yoshida16,Yoshida18}. 
Furthermore, a clear selectivity in 
the reaction region of the $^{120}$Sn$(p,p\alpha)$ was shown, meaning that only the $\alpha$ particle located on the near-side of the target nucleus with respect to the direction of the emitted $\alpha$ is observed; 
see Fig.~8 of Ref.~\cite{Yoshida16}.
It should be noted that, as emphasized in Ref.~\cite{Yoshida16}, the selectivity not only in the radius but also in the direction of the target nucleus was suggested.
We henceforth refer this to as the reaction position selectivity (RPS).
It is noted that in Ref.~\cite{Yoshida18}, the radial selectivity of ($p,p\alpha$) reaction is investigated.
When the RPS is realized, one may also expect some selectivity in spin-dependent observables. 
In fact, the effective polarization~\cite{Maris58,Jacob76,Maris79,Krein95}, 
which is sometimes called the Maris effect, 
is a well-known phenomenon in $(p,pN)$ reaction studies. 
A compact review of the Maris effect can be found in Sec.~2.3 of Ref.~\cite{Wakasa17}. 
In short, the nucleons in a nucleus in the $j_> \equiv l+1/2$ and $j_< \equiv l-1/2$ s.p.~orbits, 
where $l$ is the nucleon orbital angular momentum ($l \neq 0$), 
are effectively polarized in opposite directions. 
It is argued that a rather strong spin correlation of nucleon-nucleon (NN) scattering 
at intermediate energies and the short mean-free path of an outgoing proton 
with low energy realize the RPS and hence the Maris effect.

The purpose of this study is to clarify the role of the RPS in $(p,p\alpha)$ reactions. 
Because $\alpha$ is a spinless particle, we do not have its effective polarization. 
Nevertheless, we can still consider a polarization of the reaction residue, which is also the case with the nucleon knockout reaction. 
In this study, we consider the $^{120}{\rm Sn}(p,p\alpha){}^{116}{\rm Cd}_{2^+}$ and $^{20}$Ne($p,p\alpha$)$^{16}\mathrm{O}_{2^+}$ reactions at~392~MeV 
and investigate how the reaction residue in the $2^+$ excited state is polarized because of the RPS. 
In contrast to the previous research on the vector analyzing power of the 
($p,p\alpha$) reactions~\cite{Wang85,Neveling08,Cowley09,Mabiala09,Cowley21},
a point in this work is that the unpolarized proton beam is considered and
the effective polarization is not induced by any spin-dependent interactions.

It should be noted that in Ref.~\cite{Hansen03}, spin alignment
in one-nucleon removal reactions was discussed. The authors showed
that the parallel momentum distribution of the $^{27}$Mg$_{5/2^+}$ 
residue produced by removing a $d_{5/2}$ neutron from $^{28}$Mg
with a $^{9}$Be target is dominated by the maximum magnetic quantum
number components $m_l = \pm 2$.
The mechanism of this spin alignment in Ref.~\cite{Hansen03} is
similar to what we discuss below.
However, in the present study, we aim at clarifying that just by considering
kinematically complete measurement for $\alpha$ knockout reactions, 
without any help of the intrinsic spin of the particles involved, 
the spin {\it polarization} of the reaction residue can be achieved.

The construction of this paper is as follows. 
In Sec.~\ref{sec:framework} we describe the DWIA formalism for the ($p,p\alpha$) reaction and the treatment of the reorientation of the quantization axis. We then show in Sec.~\ref{sec:results} 
numerical results and how the residue of the ($p,p\alpha$) reaction is effectively polarized. Finally, a summary is given in Sec.~\ref{sec:summary}.

\section{THEORETICAL FRAMEWORK}\label{sec:framework}

%\subsection{Distorted wave impulse approximation}
In the present study, the factorized form of the distorted wave impulse approximation (DWIA)
without the spin degrees of freedom is employed.
This framework has been recently applied to several studies on ($p,p\alpha$) 
reactions~\cite{Yoshida16,Lyu18,Lyu19,Yoshida19,Taniguchi_48Ti}.
The kinematics of the reaction is defined following  
the Madison convention~\cite{madison_convention}. 
The reduced transition matrix $\bar{T}$ within the DWIA
framework~\cite{Wakasa17,Chant77,Chant83}
is given by
\begin{align}
\bar{T}_{m}
=&
%\sum_{\mu'_1 \mu_1  \mu'_0 \mu_0}
%\tilde{t}_{p\alpha} 
\int d\bm{R}\,
\chi_{1,\bm{K}_1}^{(-)*}(\bm{R})
\chi_{2,\bm{K}_2}^{(-)*}(\bm{R})
\chi_{0,\bm{K}_0}^{(+)}(\bm{R}) \nonumber \\
& \times e^{-i\alpha_{R}\bm{K}_0 \cdot\bm{R}}
\psi_{nlm}(\bm{R}).
\label{eq:tmatrix}
\end{align}
The incident, emitted protons, and emitted $\alpha$ are labeled as particles 0, 1, and 2, respectively,
while the bound $\alpha$ in the initial state is labeled as b.
The core nucleus in the initial state (the residue in the final state) is labeled as $\mathrm{B}$
in this paper.
$\chi_{i}$ is a distorted wave of particle $i$~($=0,1,$ or $2$) with its asymptotic momentum $\bm{K}_i$.
The third component of an angular momentum corresponds to the $z$ component unless otherwise noted.
The outgoing and incoming boundary conditions of the distorted waves 
are denoted by superscripts $(+)$ and $(-)$, respectively.
$\alpha_{R}$ is the mass ratio of the struck particle to the target, $4/120=1/30$ for $^{120}$Sn($p,p\alpha$)$^{116}{\rm Cd}_{2^+}$ and $4/20=1/5$ for $^{20}$Ne($p,p\alpha$)$^{16}\mathrm{O}_{2^+}$.
$n$ is the principal quantum number, and $l$ and $m$, respectively, are the 
orbital angular momentum and its third component of b. 
Since $\alpha$ is a spinless particle,
$\psi^n_{lm}$ is written as
\begin{align}
  \psi_{nlm}(\bm{R})
  &=
  \varphi_{nl}(R)Y_{lm}(\hat{\bm{R}}),
  \label{eq:phi-decomposition}
\end{align}
where $\varphi_{nl}$ is the radial part of the bound-state wave function and 
$Y_{lm}$ is the spherical harmonics. In the present DWIA framework, 
the triple differential cross section (TDX) with respect to the 
proton emission energy $T_1^\mathrm{L}$, its emission angles 
$\Omega_1^\mathrm{L}$, and the $\alpha$ emission angle $\Omega_2^\mathrm{L}$
is given by
\begin{eqnarray}
\frac{d^3\sigma^{\mathrm{L}}}{dT_1^\mathrm{L} d\Omega_1^\mathrm{L} d\Omega_2^\mathrm{L}}
&=&\sum_{m_y}
\left(
\frac{d^3\sigma^{\mathrm{L}}}{dT_1^\mathrm{L} d\Omega_1^\mathrm{L} d\Omega_2^\mathrm{L}}
\right)_{m_y}, 
\label{eq:tdx}\\
\left(
\frac{d^3\sigma^{\mathrm{L}}}{dT_1^\mathrm{L} d\Omega_1^\mathrm{L} d\Omega_2^\mathrm{L}}
\right)_{m_y}
&=&F_\mathrm{kin}^{\mathrm{L}} 
%\mathcal{J}_\mathrm{LG}
\frac{E_1 E_2 E_\mathrm{B}}{E_1^{\mathrm{L}} E_2^{\mathrm{L}} E_\mathrm{B}^{\mathrm{L}}}
\frac{(2\pi)^4}{\hbar v_\alpha} 
\frac{1}{2l+1} \nonumber \\
& &\times\frac{(2\pi \hbar^2)^2}{\mu_{p\alpha}}
\frac{d\sigma_{p\alpha}}{d\Omega_{p\alpha}} 
\left| \bar{T}_{m_y} \right|^2.
\label{eq:tdx_m}
\end{eqnarray}
Quantities with superscript $\mathrm{L}$ are evaluated in the laboratory frame while 
the others are in the center-of-mass frame of the three-body system.
Here, $F_\mathrm{kin}^{\mathrm{L}}$ is the phase volume
\begin{align}
F_\mathrm{kin}^{\mathrm{L}} 
=&
\frac{E_1^{\mathrm{L}} K_1^{\mathrm{L}} E_2^{\mathrm{L}} K_2^{\mathrm{L}}}{(\hbar c)^4} \nonumber \\
& \times \left[
1 + \frac{E_2^{\mathrm{L}}}{E_{\mathrm{B}}^{\mathrm{L}}}
+\frac{E_2^{\mathrm{L}}}{E_{\mathrm{B}}^{\mathrm{L}}}
\frac{
\left(
\bm{K}_1^{\mathrm{L}} - \bm{K}_0^{\mathrm{L}}
\right)
\cdot \bm{K}_2^{\mathrm{L}}
}{
\left( K_2^{\mathrm{L}} \right)^2
}
\right]^{-1},
\end{align}
and $v_\alpha$, $\mu_{p\alpha}$, $T_i$, and $E_i$ respectively, are 
the relative velocity of the incident proton and the target,
reduced energy of $p$ and $\alpha$, 
the kinetic and the total energy of particle $i$~($=0,1,2$ or $\mathrm{B}$).
$d\sigma_{p\alpha}/d\Omega_{p\alpha}$ is the $p$-$\alpha$ elastic differential cross section in free space with $p$-$\alpha$ two-body scattering energy and angle determined by the ($p,p\alpha$) kinematics.
It should be noted that the TDX is decomposed into the components of each $m_y$, 
which is the projection of $l$ on the $y$-axis 
taken to be the direction of $\bm K_0 \times \bm K_1$.
In other words, in the calculation of the TDX of Eqs.~(\ref{eq:tdx}) and (\ref{eq:tdx_m}), 
the $y$-axis is taken as the quantization axis. 
See Sec. 3.1 of Ref.~\cite{Wakasa17} for details.

% In Sec.~\ref{sec:result}
The transition matrix $\bar{T}_{m_y}$ in Eq.~(\ref{eq:tdx_m}) can be obtained by 
reorientating the quantization axis of $\bm{l}$ 
from the $z$-axis to the $y$-axis,
\begin{align}
  \bar{T}_{m_y}
  &=
  \sum_{m}D^{l}_{m_y,m}(\hat{\bm{R}}_{yz})
  \bar{T}_m,
  \label{eq:T-conversion}
\end{align}
where $D$ is the Wigner's $D$-matrix and $\hat{\bm{R}}_{yz}$ is a
rotation represented by Euler angles from the $z$-axis to the $y$-axis.
The orthonormality and completeness of $D$ ensure the invariance of the TDX
under the reorientation of the quantization axis,
\begin{align}
  \sum_{m_y}\left|\bar{T}_{m_y}\right|^2
  &=
  \sum_{m  }\left|\bar{T}_{m  }\right|^2.
\end{align}
In the present study, $D^{l}_{m_y,m}$ in Eq.~(\ref{eq:T-conversion}) only acts on 
the spherical harmonics $Y_{lm}$ in Eq.~(\ref{eq:phi-decomposition})
and reorientates its quantization axis.

\section{RESULTS AND DISCUSSION}\label{sec:results}
\subsection{Numerical input}
\begin{figure}[h]
  \centering
  \includegraphics[width=0.7\hsize]{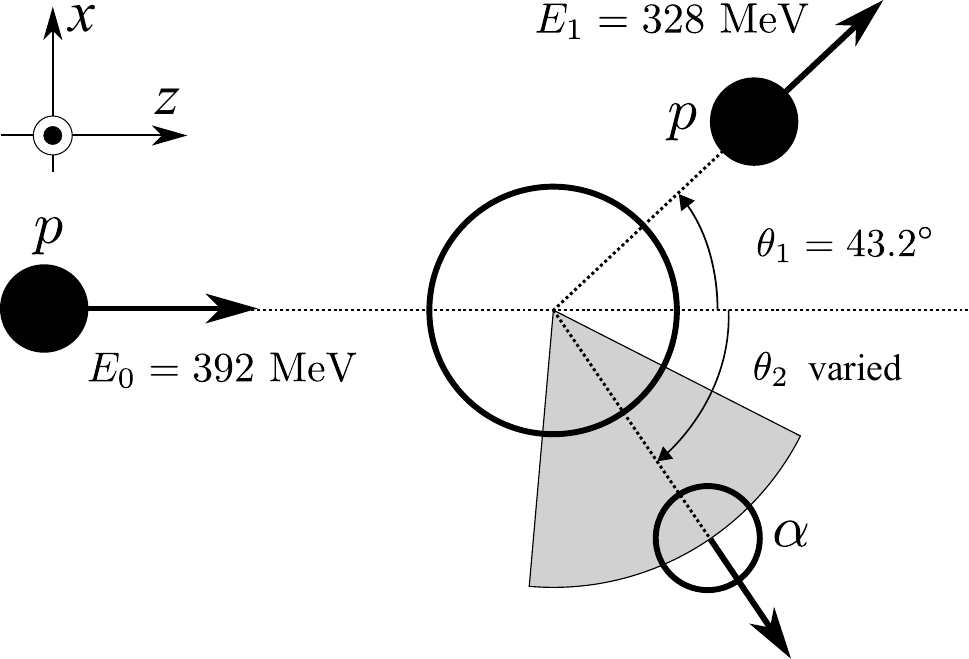}
  \caption{The kinematical setup of the $^{120}$Sn($p,p\alpha$)$^{116}\mathrm{Cd}_{2^+}$ reaction.}
  \label{fig:kinematics}
\end{figure}

First, we consider the $^{120}$Sn($p,p\alpha$)$^{116}\mathrm{Cd}_{2^+}$ cross section at 392~MeV.
The recoilless condition, in which the momentum of the residue B in the
final state is zero ($\bm{K}^\mathrm{L}_{\mathrm{B}} = 0$),
is realized when $T^\mathrm{L}_1 = 328$~MeV, $\theta_1^\mathrm{L} = 43.2^\circ$, 
$\phi_1^\mathrm{L} = 0^\circ$, 
$\theta_2^\mathrm{L} = 61^\circ$, and $\phi_2^\mathrm{L} = 180^\circ$.
These conditions are determined by an available setup at RCNP, expecting such experiments will be conducted in the near future.
The TDX$_{m_y}$ around the recoilless condition is calculated by 
varying $\theta_2^\mathrm{L}$.
The kinematical setup is shown in Fig.~\ref{fig:kinematics}.

The $\alpha$ cluster wave function $\psi_{nlm}$ is obtained 
as a bound state of $\alpha$ and $^{116}$Cd within a Woods-Saxon 
potential,
\begin{align}
  V(R) &=
  \frac{V_0}{1+\mathrm{exp}\left(\frac{R-R_0}{a}\right)}.
\end{align}
Its range and diffuseness parameters are $R_0 = 1.27 \times 116^{1/3}$~fm and $a = 0.67$~fm, respectively.
The depth parameter $V_0$ is determined to reproduce the $\alpha$ separation energy of $^{120}$Sn.
We consider $n = 7$ and $l = 2$, and $\psi_{nlm}$ is normalized to unity; 
$n$ is determined by the orthogonality condition model~\cite{Saito1969}.
% In Fig.\ref{fig;rwf}, the solid, dotted, and dashed lines correspond to $\varphi_{72}(R), R\varphi_{72}(R)$, and $R^2\varphi_{72}(R)$, respectively.
% The integrand of Eq.\eqref{eq:tmatrix} in spherical coordinates is separated into the product of the functions with angles as arguments and $R^2\varphi_{72}(R)$. Now $R^2\varphi_{72}(R)$ is regarded as the radial distribution of the $\alpha$ bound state, and
% it has an amplitude in the nuclear interior region
% ($R \lesssim 6$~fm), which does not contribute to the TDX due to the absorption effect~\cite{Yoshida16,Yoshida18}.
The $p$-$\alpha$ differential cross section in free space 
in Eq.~(\ref{eq:tdx_m}) is obtained by the microscopic single-folding 
model~\cite{Toyokawa13} with a
phenomenological $\alpha$ density in free space and the Melbourne nucleon-nucleon \textit{g}-matrix interaction~\cite{Amos00}.
For the $\alpha$ density, we use 
the phenomenological proton density~\cite{Vries1987} determined from 
electron scattering in which the finite-size effect 
due to the proton charge is unfolded in the standard manner~\cite{Singhal1978}.
The neutron density is assumed to have the same geometry as the proton one. 
The $p$-$^{120}$Sn and $p$-$^{116}$Cd distorted waves are obtained 
as a scattering state under the Koning-Delaroche 
global optical potential~\cite{Koning03}.
As for the $\alpha$-$^{116}$Cd distorted wave, the global $\alpha$ optical 
potential proposed by Avrigeanu \textit{et al.}~\cite{Avrigeanu94}
is adopted.
For comparison, we also use microscopic optical potentials. The $p$-$^{120}$Sn, $p$-$^{116}$Cd, and $n$-$^{116}$Cd potentials are obtained by the single-folding model with the Melbourne $g$-matrix and nuclear densities of $^{120}$Sn and $^{116}$Cd calculated with the Bohr-Mottelson s.p. potential~\cite{Bohr69}. The $\alpha$-$^{116}$Cd potential is obtained by the nucleon–nucleus folding (NAF) model~\cite{Egashira14} with the $p$-$^{116}$Cd and $n$-$^{116}$Cd potentials.

% \begin{figure}[h]
%   \centering
%   \includegraphics[width=1.0\hsize]{ffr_120Sn.pdf}
%   \caption{Radial part of $\alpha$ bound-state wave function. 
%   The solid, dotted, and dashed lines correspond to $\varphi_{72}(R),
%   R\varphi_{72}(R)$, and $R^2\varphi_{72}(R)$, respectively.}
%   \label{fig;rwf}
% \end{figure}

\subsection{Effective polarization in ($p,p\alpha$) reaction}
\begin{figure}[h]
  \centering
  \includegraphics[width=1.0\hsize]{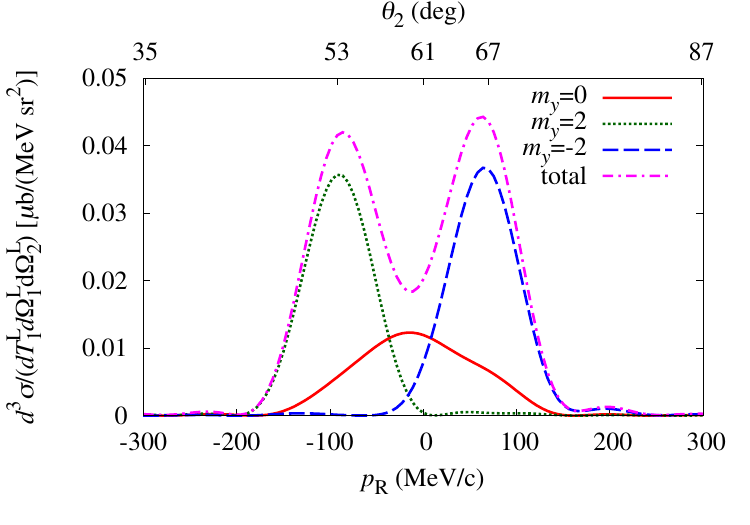}
  \caption{TDX$_{m_y}$ of $^{120}$Sn($p,p\alpha$)$^{116}$Cd$_{2^+}$ at 392~MeV 
  as a function of the recoil momentum.
  The solid, dotted, and dashed lines represent the components for 
  $m_y=0,2$, and $-2$, respectively. The dot-dashed line shows 
  the sum of all the components.}
  \label{fig:TDX_DWIA}
\end{figure}

\begin{figure}[h]
  \centering
  \includegraphics[width=0.5\hsize]{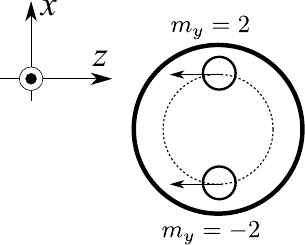}
  \caption{
   Classical explanation of the $\alpha$ motion inside the target 
   when $p_\mathrm{R}$ is positive. 
   In this case, $\bm{K}^\mathrm{L}_\alpha$ is parallel to the $-z$ direction. 
   Because the $\alpha$ particle with $m_y=2$ ($-2$) revolves counterclockwise 
   (clockwise), its position is specified as in the figure.
  }
  \label{fig:classical}
\end{figure}

We show in Fig.~\ref{fig:TDX_DWIA} the TDX$_{m_y}$ 
as a function of $p_\mathrm{R}$:
\begin{equation}
  p_{\mathrm{R}} 
  =
  \hbar K^\mathrm{L}_\mathrm{B}
  \frac{K^\mathrm{L}_{\mathrm{B}z}}{\left|K^\mathrm{L}_{\mathrm{B}z}\right|},
\end{equation}
with $K^\mathrm{L}_{\mathrm{B}z}$ being the $z$ component of 
$\bm{K}^\mathrm{L}_{\mathrm{B}}$.
It is clearly seen that $m_y = 2$ and $m_y = -2$ components are
well selected by the kinematics.
This is explained as follows.
Considering the quasi-free knockout reaction, momentum $\bm{K}^\mathrm{L}_\alpha$ 
of b~(the bound $\alpha$ particle in A) 
has an opposite momentum to $\bm{K}_\mathrm{B}^\mathrm{L}$,
since the target is at rest in the laboratory frame. 
Thus, depending on $m_y$, the position of b to be knocked out can be specified. 
This implies that b {\lq\lq}moves'' to the left (the $-z$ direction) when $p_\mathrm{R}$ 
is positive as shown in Fig.~\ref{fig:classical}.

Considering the kinematics that
$\alpha$ is emitted to $\theta_2^\mathrm{L} = 66.5^\circ$ and $\phi_2^\mathrm{L} = 180^\circ$, 
the $m_y = 2$ component in Fig.~\ref{fig:classical} cannot be knocked out because
of the short mean-free path of the $\alpha$ particle, which is described  
by the absorption of the $\alpha$-$^{116}$Cd optical potential
in the present framework.
On the other hand, the $m_y = -2$ component is almost free from the absorption.
Therefore the total TDX is dominated by the $m_y = -2$ component when $p_\mathrm{R}$ is
positive.  Similarly, when $p_\mathrm{R}$ is negative, the $m_y=2$ component becomes dominant.
This mechanism has similarities and differences to 
the Maris effect~\cite{Maris58,Jacob76,Maris79,Krein95} 
in the nucleon knockout reactions.
As for similarities, a classical picture of the orbital motion of the bound particle works to explain the phenomena. 
In both the ($p$,$pN$) and the ($p$,$p\alpha$) cases, the reaction position is restricted to only two positions once the momentum of the bound particle, that is, $\bm{K}_{\alpha}^{\mathrm{L}}$ in the present study, is given.
In addition, the strong absorption effect excludes the contribution from the ``far side'' of the two reaction positions with respect to the emitted $\alpha$ direction.
Regarding differences, 
in the Maris effect for ($p,pN$), the energy of one emitted nucleon is low and that of the other is high, and the former is absorbed. 
In contrast, in ($p$,$p\alpha$), the RPS is realized by the strong absorption, that is, the short mean-free path, of the $\alpha$ particle compared to that of a nucleon.
Another typical difference is that in the vector analyzing power of ($p$,$pN$), the Maris effect distinguishes the $j_{>}$ and $j_{<}$ orbitals for a given angular momentum $l$ by utilizing the strong spin correlation of the NN system.
On the other hand, in ($p$,$p\alpha$), the present mechanism discriminates $m_y$ and therefore the spin third component of the residue only by the kinematical condition.
There is no need for a polarized beam because the transition involves the spinless $\alpha$ particle and the reaction is independent of the spin degrees of freedom. 
% A unique feature in this case is that the selection on $m_y$ and hence 
% a selection on the spin third component of the residue is accomplished 
% by only a condition on the kinematics and the absorption effect, without spin degrees of freedom.

\begin{figure}[h]
  \centering
  \includegraphics[width=1.0\hsize]{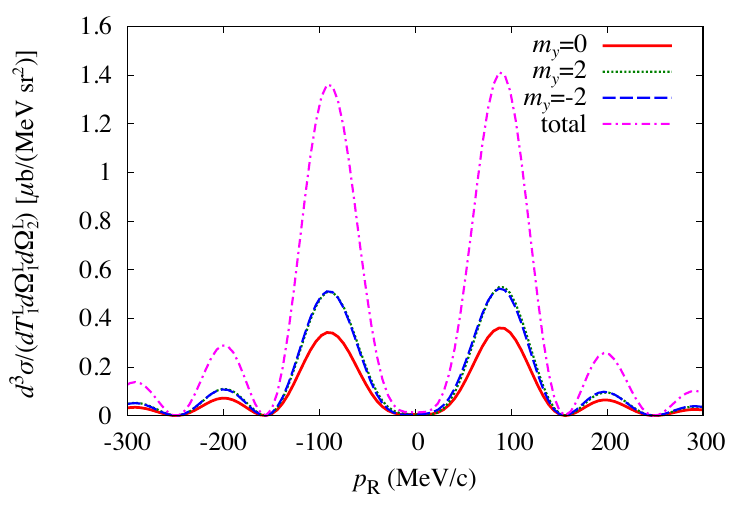}
  \caption{The same as Fig. \ref{fig:TDX_DWIA} but with the plane-wave limit.}
  \label{fig:TDX_PWIA}
\end{figure}

It should be noted that the TDX$_{m_y=\pm 1}$ vanishes because the integrand in 
Eq.~(\ref{eq:tmatrix}) is anti-symmetric with respect to the $z$-$x$ plane 
if $l-m$ is odd.  
This property originates from that of $Y_{lm_y}$; the remaining part 
of the integrand in Eq.~(\ref{eq:tmatrix}) is symmetric with respect to 
the $z$-$x$ plane because the kinematics are taken in coplanar on that plane. 
Figure~\ref{fig:TDX_PWIA} shows the same result as in Fig.~\ref{fig:TDX_DWIA} 
but with the plane-wave limit.
The selectivity found in the DWIA calculation completely disappears and 
this result clearly shows that the $m_y$ selection is achieved by 
the absorption effect on the $\alpha$ particle.

\begin{figure}[h]
  \centering
  \includegraphics[width=1.0\hsize]{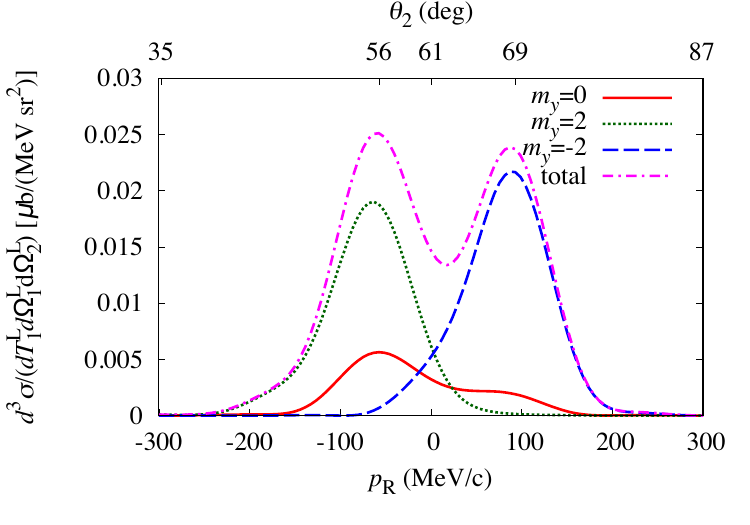}
  \caption{The Same as Fig.~\ref{fig:TDX_DWIA} but with microscopic optical potentials.}
  \label{fig:TDX_DWIA_micro}
\end{figure}

Figure~\ref{fig:TDX_DWIA_micro} displays the results shown in Fig.~\ref{fig:TDX_DWIA} but calculated with microscopic optical potentials.
As seen, the absolute values of each component of the TDX and its peak positions are changed from those in Fig.~\ref{fig:TDX_DWIA}. For the $m_y=0$ component, a difference in the shape is also observed. Nevertheless, the dominance of the $m_y=2$ ($-2$) component around the peak with negative (positive) $p_{\mathrm{R}}$ is found to be robust. For more quantitative discussion, a direct comparison between the results with DWIA and experimental data will be necessary. Investigation of the role of higher-order processes that are not included in DWIA will also be very important. For this purpose, a reaction model (theory) that is applicable to the TDX calculation in complete kinematics, e.g., the FAGS theory, will be needed. 

\begin{figure}[h]
  \centering
  \includegraphics[width=1.0\hsize]{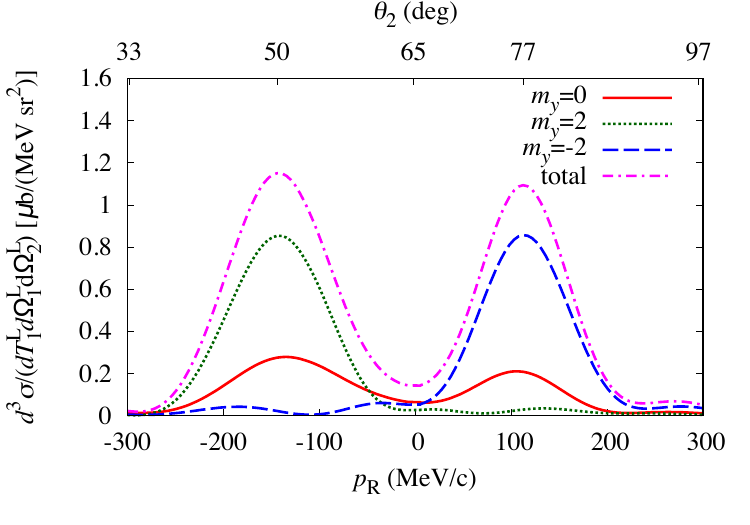}
  \caption{TDX$_{m_y}$ of $^{20}$Ne($p,p\alpha$)$^{16}$O$_{2^+}$ at 392~MeV 
  as a function of the recoil momentum. 
  The same manner as Fig. \ref{fig:TDX_DWIA}.}
  \label{fig:TDX_Ne}
\end{figure}

We also calculate the $^{20}$Ne($p,p\alpha$)$^{16}$O$_{2^+}$ cross section at 392~MeV.
The recoilless condition is realized when $T^\mathrm{L}_1 = 341$~MeV, $\theta_1^\mathrm{L} = 35^\circ$, 
$\phi_1^\mathrm{L} = 0^\circ$, 
$\theta_2^\mathrm{L} = 65^\circ$, and $\phi_2^\mathrm{L} = 180^\circ$.
One sees that the shape of the $m_y=0$ component (solid line) considerably differs from that in Fig.~\ref{fig:TDX_DWIA} and is similar to what we expect in the PW limit. Thus, this result implies that the distortion effect is weak compared to the $^{120}$Sn($p,p\alpha$)$^{116}$Cd$_{2^+}$ case. Nevertheless, the $m_y$ selection still appears in Fig.~\ref{fig:TDX_Ne}, which will indicate that the nuclear absorption plays a role also in the $^{20}$Ne($p,p\alpha$)$^{16}$O$_{2^+}$ case.
These results strongly suggest that the effective polarization by the ($p,p\alpha$) reaction is universal from light to heavy nuclei. 

\section{SUMMARY}\label{sec:summary}
We have demonstrated that an effective polarization of 
the residual nucleus is realized in 
the $^{120}$Sn($p,p\alpha$)$^{116}$Cd$_{2^+}$ and $^{20}$Ne($p,p\alpha$)$^{16}$O$_{2^+}$ reaction at 392~MeV. 
The mechanism of this effective polarization is similar to 
that of the Maris effect but as a distinct feature, 
the effective polarization in the ($p,p\alpha$) processes has nothing to do with 
the spin degrees of freedom of the reacting particles. 
In other words, by just measuring the TDX with slightly varying kinematics, 
a polarized residual nucleus, $^{116}$Cd in the $2_1^+$ state in this case, 
can be extracted. 
This scenario is considered to be the same as for the residue in other spin states.
This may be a useful polarization technique based on 
a kinematically complete direct reaction 
in both normal and inverse kinematics.

\section*{ACKNOWLEDGMENTS}
We thank T.~Uesaka for fruitful discussions. 
This work is supported in part by Grants-in-Aid for Scientific Research 
from the JSPS (Grants No.~JP20K14475, No.~JP21H00125, and No.~JP21H04975) and by JST SPRING, Grant No. JPMJSP2138.

% \nocite{*}

\bibliography{mybibfile}% Produces the bibliography via BibTeX.

\end{document}